\documentclass{article}
\usepackage{amssymb}
\usepackage{amsmath}
\usepackage{latexsym}
\usepackage{array}
\usepackage{graphicx}
\usepackage{bm}
\usepackage{exscale}

\usepackage{theorem}
\theoremstyle{break}
\newtheorem{Theorem}{Theorem}%[section]
\newtheorem{Proposition}[Theorem]{Proposition}%[section]
\newtheorem{Lemma}[Theorem]{Lemma}
\newtheorem{Definition}[Theorem]{Definition}%{section}
\newtheorem{Corollary}[Theorem]{Corollary}
\def\qed{\hfill\hbox{$\Box$}\vspace{10pt}\break}

\begin{document}
\title{Integrability criterion in terms of coprime property for the discrete Toda equation}
\author{Masataka Kanki$^1$, Jun Mada$^2$ and Tetsuji Tokihiro$^3$\\
\small $^1$ Department of Mathematics, Faculty of Science,\\
\small Rikkyo University, 3-34-1 Nishi-Ikebukuro, Tokyo 171-8501, Japan\\
\small $^2$ College of Industrial Technology,\\
\small Nihon University, 2-11-1 Shin-ei, Narashino, Chiba 275-8576, Japan\\
\small $^3$ Graduate School of Mathematical Sciences,\\
\small University of Tokyo, 3-8-1 Komaba, Tokyo 153-8914, Japan
}
%\date{}
\maketitle

\begin{abstract}
We reformulate the singularity confinement, which is one of the most famous integrability criteria for discrete equations, in terms of the algebraic properties of the general terms of the discrete Toda equation.
We show that the coprime property, which has been introduced in our previous paper
as one of the integrability criteria, is appropriately formulated and proved for the discrete Toda equation.
We study three types of boundary conditions (semi-infinite, molecule, periodic) for the discrete Toda equation, and prove that the same coprime property holds for all the types of boundaries.

\texttt{MSC2010: 37K10, 35A20, 47A07}

\texttt{Keywords: integrable system, discrete Toda equation, singularity confinement}
\end{abstract}

%%%%%%%%%%%%%%%%%%%%%%%%%%%%%%%%%%%%%%%%%%%%%%%%%%
\section{Introduction}
\label{sec1}
Continuous integrable systems are nonlinear differential equations that can be solved analytically. For example, integrability of ordinary differential equations (ODEs) is judged by the Arnold-Liouville theorem, which requires integrable
ODEs to have sufficient number of first integrals (i.e., conserved quantities, constants of motions) \cite{Arnold}.
There is little ambiguity in the notion of integrability in continuous cases.
On the other hand, in the case of discrete equations, universally accepted definition of integrability does not exist.
One of the most widely used criteria for integrability might be the `singularity confinement test' (SC test) introduced in \cite{SC} by B. Grammaticos, A. Ramani and V. Papageorgiou, as a discrete analogue of the Painlev\'{e} property \cite{Conte}.
According to the SC test, a difference equation is considered to be integrable, if every singularity of the equation is cancelled out to give a finite value after a finite number of iterations of the mapping.
The SC test has been successfully applied to several types of ordinary difference equations, in particular the non-autonomous generalizations of the QRT mappings \cite{QRT}, to produce discrete versions of the Painlev\'{e} equations \cite{RGH}.
On the other hand, it is usually not easy to conduct the SC test to partial difference equations.
Indeed, there is a result on the SC test of partial difference equations in their bilinear forms \cite{RGS}, where the Hirota-Miwa equation and its reductions are studied. Also, the singularity confinement of the discrete KdV equation in its nonlinear form is discovered in \cite{SC}, where two patterns of confining singularities
on the lattice are presented.
However, in both cases, not all the patterns of singularities have been investigated.
One of the most difficult points in conducting the SC test for partial difference equations is that, it is not practical to investigate whether all the patterns of singularities are eliminated after finite number of iterations of the given equation, because the partial difference equations have infinite dimensional (or high dimensional depending on the size of the system) space of initial conditions.
To overcome this problem, we have introduced in our previous papers a method to reformulate the SC test in terms of the algebraic relations of the general terms of the equations \cite{dKdVSC,dKdVSC2}.
In these papers, we have introduced the notion of `co-primeness', which can be used as a new integrability criterion for both ordinary and partial difference equations, and have proved the co-primeness theorems for a type of QRT mappings and the nonlinear form of the discrete KdV equation.
In these previous works, we treated those equations under the semi-infinite boundary conditions. In the proof of the co-primeness, we utilized the fact that the bilinear forms of those equations have the Laurent property, which has already been established in relation with the notion of cluster algebras \cite{FZ,FZ2}.
In the case of Dirichlet and periodic boundary conditions, however, the Laurent property of integrable equations has not been clarified. In fact, as we shall see below in corollary \ref{periodthm2}, the Laurent property does not hold in its naive form for the periodic boundary condition.
%
%space

%
The aim of this paper is to examine whether  co-primeness theorems similar to those in our previous works are satisfied
for integrable equations with boundary conditions other than the semi-infinite one.
For this purpose, we consider the celebrated discrete Toda equation under three types of boundary conditions: i.e., semi-infinite, molecule, and periodic.
We shall prove that the co-primeness theorem does hold for these boundary conditions.
The Toda lattice equation has been introduced by M. Toda as a mechanical model of the chain of particles under nonlinear interaction force \cite{Toda1}.
It is an important example of integrable systems with multi-soliton solutions. It reduces to the KdV equation with an appropriate continuum limit \cite{Toda2}.
The Toda equation has numerous applications to physical phenomena, such as a wave propagation on two-dimensional water surfaces, an electric current in circuits.
Later the time discretization of the Toda equation has been studied and it has been shown that the system is completely integrable \cite{Suris1, Suris2}.
The discrete Toda equation is the following coupled equations:
\begin{eqnarray}
I_n^{t+1} &=& I_n^t+V_n^t-V_{n-1}^{t+1},\label{dtodaIV1} \\
V_n^{t+1} &=& \frac{I_{n+1}^t V_n^t}{I_n^{t+1}}, \label{dtodaIV2}
\end{eqnarray}
with suitable boundary conditions. For example, in the case of semi-infinite boundary condition, we take $V_0^t=0$ for $t\ge 0$.
In the case of molecule boundary condition, we take $V_0^t=V_{N+1}^t=0$ for $t\ge 0$,
where $N$ is the size of the system.
In the case of periodic boundary condition, we take
$V_n^t=V_{N+n}^t$, $I_n^t=I_{N+n}^t$ for $t\ge 0$ and $n\ge 0$.
\begin{figure}
\centering
\includegraphics[width=11cm,bb=70 200 750 550]{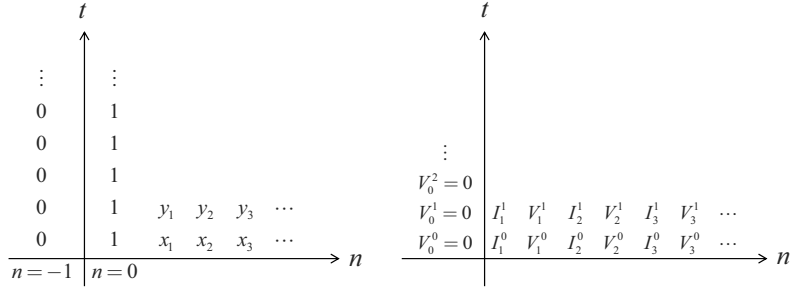}
\caption{Initial values of discrete Toda equation with semi-infinite boundary condition, where $x_i:=\tau_i^0$, $y_j:=\tau_j^1$.}
\label{figure1}
\end{figure}
\begin{figure}
\centering
\includegraphics[width=11cm,bb=70 280 750 550]{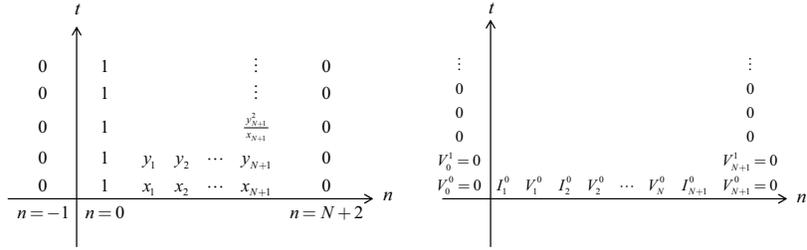}
\caption{Initial values of discrete Toda equation with molecule boundary condition, where $x_i:=\tau_i^0$, $y_j:=\tau_j^1$.}
\label{figure2}
\end{figure}
The bilinear form of the discrete Toda equation is as follows:
\begin{equation}
\tau_n^{t+1} \tau_n^{t-1} = \tau_{n-1}^{t+1} \tau_{n+1}^{t-1} + (\tau_n^t)^2. \label{dtoda}
\end{equation}
The boundary condition for the equation \eqref{dtoda} is determined in accordance with that of equations \eqref{dtodaIV1}, \eqref{dtodaIV2}.
The following proposition \ref{prop1} determines the correspondence between two sets of equations for the molecule boundary condition. The correspondence for the semi-infinite boundary can be obtained with the limit $N\to \infty$. The case of periodic boundary condition is discussed later in section \ref{sec2}.
\begin{Proposition} \label{prop1}
If we are given the solution $\tau_n^t$ of \eqref{dtoda} with conditions
$\tau_{-1}^t=\tau_{N+2}^t=0$ $(t\ge 0)$,
then the following set of variables $I_n^t$ and $V_n^t$ defined by \eqref{tauIV} satisfy the discrete Toda molecule equation (i.e., \eqref{dtodaIV1} and \eqref{dtodaIV2} with conditions $V_0^t=V_{N+1}^t=0$ for all $t\ge 0$):
\begin{equation}\label{tauIV}
I_n^t=\frac{\tau_{n-1}^t \tau_n^{t+1}}{\tau_n^t \tau_{n-1}^{t+1}},\;\; V_n^t=\frac{\tau_{n+1}^t \tau_{n-1}^{t+1}}{\tau_n^t \tau_n^{t+1}}.
\end{equation}
\end{Proposition}
See figures \ref{figure1} and \ref{figure2} for the configurations of initial values.
Let us fix the definition of co-primeness of Laurent polynomials and rational functions here.
\begin{Definition} \label{laurentcoprime}
Two Laurent polynomials $f,g$ are `co-prime' in the ring $R:=\mathbb{Z}[a_i^{\pm}; 1\le i\le n]$ if the following condition is satisfied:
If we have decompositions $f=hf_2$, $g=hg_2$ in $R$, then $h$ must be a unit in $R$ (i.e., a monomial in $\{a_i\}_{i=1}^n$ with coefficient $1$).
\end{Definition}
\begin{Definition} \label{rationalcoprime}
Two rational functions $f$ and $g$ are `co-prime' in the field $F:=\mathbb{C}(a_i; 1\le i\le n)$ if the following condition is satisfied:
Let us express $f,g$ as $f={F_1}/{F_2}$ and $g={G_1}/{G_2}$
where $F_i,G_i\in \mathbb{C}[a_i^{\pm}; 1\le i\le n]$ $(i=1,2)$, $(F_1,F_2)$ and $(G_1,G_2)$ are coprime pairs of polynomials.
Then every pair of polynomials $(F_i,G_j)$ $(i,j=1,2)$ is coprime in the sense of definition \ref{laurentcoprime}. (No common factor except for monomial one is allowed.)
\end{Definition}

\begin{Lemma}[\cite{dKdVSC2}] \label{locallemma}
Let $\{p_1,p_2,\cdots,p_m\}$ and $\{q_1,q_2,\cdots ,q_m\}$ be two sets of independent variables with the following properties:
\begin{eqnarray}
p_j&\in&\mathbb{Z}\left[ q_1^{\pm}, q_2^{\pm},\cdots ,q_m^{\pm}\right], \label{p}\\
q_j&\in&\mathbb{Z}\left[ p_1^{\pm}, p_2^{\pm},\cdots ,p_m^{\pm}\right], \label{q}\\
q_j&&\mbox{is irreducible as an element of}\ \mathbb{Z}\left[ p_1^{\pm}, p_2^{\pm},\cdots ,p_m^{\pm}\right], \notag
\end{eqnarray}
for $j=1,2,\cdots, m$.
Let us take an irreducible Laurent polynomial
\[
f(p_1,\cdots,p_m)\in \mathbb{Z}\left[ p_1^{\pm}, p_2^{\pm},\cdots ,p_m^{\pm}\right],
\]
and another Laurent polynomial
\[
g(q_1,\cdots, q_m) \in \mathbb{Z}\left[ q_1^{\pm}, q_2^{\pm},\cdots ,q_m^{\pm}\right],
\]
which satisfies $f(p_1,\cdots,p_m)=g(q_1\cdots, q_m)$.
In these settings, the function $g$ is decomposed as
\[
g(q_1,\cdots, q_m)=p_1^{r_1}p_2^{r_2}\cdots p_m^{r_m}\cdot \tilde{g}(q_1,\cdots,q_m),
\]
where $r_1,r_2, \cdots, r_m\in\mathbb{Z}$ and $\tilde{g}(q_1,\cdots,q_m)$ is irreducible in $\mathbb{Z} \left[ q_1^{\pm}, q_2^{\pm},\cdots ,q_m^{\pm}\right]$.
\end{Lemma}
The proof can be found in \cite{dKdVSC2}.
%%%%%%%%%%%
%%%%%%%%%%%%
%

\section{Co-prime property of the discrete Toda} \label{sec2}
\subsection{Semi-infinite boundary}
\begin{figure}
\centering
\includegraphics[width=7cm,bb=100 600 300 740]{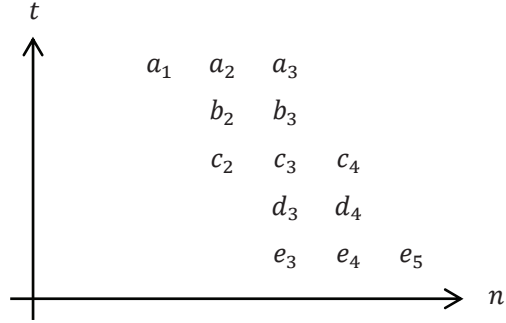}
\caption{Configuration of the variables in lemma \ref{c3a3}.}
\label{figure3}
\end{figure}

We take the initial values as
\begin{equation} \label{semiinfcond}
\tau_0^t=1\ (t=0,1,2,\cdots),\; \tau_n^0=x_n,\ \tau_n^1=y_n\ (n=1,2,\cdots).
\end{equation}
Note that taking $\tau_{-1}^t=0$ $(t\ge 0)$ together with $\tau_0^0=\tau_0^1=1$ is equivalent to imposing $\tau_0^t=1$ $(t\ge 0)$.
\begin{Theorem} \label{thmsemiinf}
Every term of the discrete Toda equation \eqref{dtoda} is a Laurent polynomial of the initial variables:
\[
\tau_n^t\in\mathbb{Z}[x_1^{\pm}, x_2^{\pm},\cdots, y_1^{\pm}, y_2^{\pm},\cdots],
\]
where $\tau_n^0=x_n$ and $\tau_n^1=y_n$.
Moreover, the term $\tau_n^t$ is an irreducible Laurent polynomial,
and two distinct terms are coprime as Laurent polynomials.
\end{Theorem}
\textbf{Proof of theorem \ref{thmsemiinf}}\;\;
Let us define the ring of Laurent polynomials as
\[
R_{m,n}:=\mathbb{Z}[x_1^{\pm},x_2^{\pm},\cdots,x_m^{\pm};y_1^{\pm},y_2^{\pm},\cdots,y_n^{\pm}],
\]
and use the notation as $R:=\lim_{m,n\to \infty} R_{m,n}$.
The subset of irreducible Laurent polynomials is
\[
R_{irr}:=\left\{f\in R| f \ \mbox{is  an irreducible element of} \ R \right\}.
\]
The following lemma is immediate:
\begin{Lemma} \label{fgirred}
For $f=a x_m^{\pm 1}+b$, $a,b\in R_{m-1,n}\setminus \{0\}$,
$f$ is irreducible in $R$ if $a$ and $b$ are coprime in $R$.
For $g= cy_n^{\pm 1}+d$, $c,d\in R_{m,n-1}\setminus \{0\}$,
$g$ is irreducible in $R$ if $c$ and $d$ are coprime in $R$.
\end{Lemma}
\begin{Lemma} \label{c3a3}
Let us rewrite
\begin{align*}
&(\tau_{n-2}^{t+2},\tau_{n-1}^{t+2},\tau_n^{t+2},\tau_{n-1}^{t+1},\tau_n^{t+1},\tau_{n-1}^{t},\tau_n^t,\tau_{n+1}^t)=(a_1,a_2,a_3,b_2,b_3,c_2,c_3,c_4),\\
&(\tau_n^{t-1},\tau_{n+1}^{t-1},\tau_n^{t-2},\tau_{n+1}^{t-2},\tau_{n+2}^{t-2})=(d_3,d_4,e_3,e_4,e_5).
\end{align*}
Then we have
\[
c_3 a_3=c_3\frac{(a_1c_3e_5+ a_1 d_4^2+ b_2^2 e_5)d_3^2+(c_3^3+2b_2c_3d_4)c_2e_4+b_2^2d_4^2e_3}{c_2 d_3^2 e_4}.
\]
See figure \ref{figure3} for the configuration of these values.
\end{Lemma}
\textbf{Proof of lemma \ref{c3a3}}\;\;
Equation \eqref{dtoda} shows that $a_3 c_3=b_3^2+a_2 c_4$.
We substitute other relations $a_2={(b_2^2+a_1 c_3)}/{c_2}$ and $c_4={(d_4^2+c_3e_5)}/{e_4}$ to the equality above. Direct calculation shows that
\[
c_3a_3=\frac{1}{c_2d_3^2e_4}\{c_3(a_1d_3^2d_4^2+c_2c_3^3e_4+2b_2c_2c_3d_4e_4+b_2^2d_3^2e_5+a_1c_3d_3^2e_5)+(b_2^2 d_4^2)(d_3^2+c_2e_4)\}.
\]
We then use $d_3^2+c_2e_4=c_3e_3$ to obtain the desired result.
\qed
Using these lemmas we prove theorem \ref{thmsemiinf}, which is restated as follows: ``We have $\tau_n^t\in R_{irr}$ for every $t,n\ge 0$, and
two distinct terms are coprime.''
We prove this by induction with respect to $t$.
\paragraph{The case of $t=2$:}
Let us rewrite $z_n:=\tau_n^2$ for simplicity.
We prove that $z_n\in R_{n+1,n}$ and that $z_n$ is an irreducible linear function with respect to $x_{n+1}$.
First, $z_1={(y_1^2+x_2)}/{x_1}\in R_{2,1}$ is trivially irreducible
(or we can use lemma \ref{fgirred}).
Since $z_1$ is not a monomial, it is coprime with $\tau_n^1 (n\ge 1)$ and $\tau_n^2 (n\ge 2)$. Next
\[
z_n=\frac{1}{x_n}(z_{n-1}x_{n+1} + y_n^2)
\]
for $n \ge 2$ tells us inductively that $z_n\in R_{n+1,n}$ and that $z_n$ is a linear function of $x_{n+1}$. Moreover $z_n$ is not a monomial because $y_n\neq 0$.
Therefore, from lemma \ref{fgirred}, we obtain inductively that $z_n$ is an irreducible Laurent polynomial.
We also have that $z_n$ and $z_m$ with $n \neq m$ are coprime,
since both $z_n$ and $z_m$ are irreducible and each one is linear with respect to $x_{n+1}$ (resp. $x_{m+1}$). It is clear that $z_n$ is coprime with $x_m$ and $y_k$ for all $n,m,k$.
\paragraph{The case of $t=3$:}
We can prove the following relation by induction:
\begin{equation} \label{yzu}
\tau_n^{t+2}=\tau_{n+1}^t \sum_{k=0}^n \frac{(\tau_k^{t+1})^2}{\tau_k^t \tau_{k+1}^{t}}.
\end{equation}
Let us rewrite $u_n:=\tau_n^3$.
By taking $t=1$ in \eqref{yzu}, we have
\[
u_n=y_{n+1}\sum_{k=0}^{n} \frac{(z_k)^2}{y_k y_{k+1}}\in R.
\]
In particular, we have $u_1={(y_2+z_1^2)}/{y_1}$. Since $z_1\in R_{2,1}$, from lemma \ref{fgirred}, the term $u_1$ is irreducible in $R$.
As $z_k$ is irreducible and is not linear in $y_2$, $z_k$ is coprime with $u_1$ for all $k$.
We next prove by induction that
\[
u_n=\frac{u_{n-1} y_{n+1}+z_n^2}{y_n}
\]
is irreducible and coprime with other elements ($\tau_n^t;$ $t=0,1,2$). Let us suppose that $u_{n-1}$ is irreducible and is coprime with $z_n$, then, using lemma \ref{fgirred}, we conclude that $u_n$ is irreducible.
Since neither  $z_j (j\ge 1)$ nor $u_k (k\le n-1)$ contains $y_{n+1}$,
while $u_n$ is linear in $y_{n+1}$,
$u_n$ is coprime with $z_j$ and $u_k$.
The proof is finished for $t=3$.
\paragraph{The case of $t\ge 4$:}
Let us define a region $\mathcal{D}_k$ in $(n,t)$-plane as
\[
\mathcal{D}_k=\{(n,t)\, |\, 1\le n \le k,\, 0\le t\le 2k-2n+1 \},
\]
where $k=1,2,\cdots$,
and prove that theorem \ref{thmsemiinf} is true in the region $\mathcal{D}_k$ by induction.
The case of $k=1$ is trivial because $\mathcal{D}_1=\{(1,0),(1,1)\}$.
The case of $k=2$ is true from the previous two paragraphs for $t=2,3$,
since $(n,t)\in \mathcal{D}_2$ always satisfies $t\le 3$.
Let us assume that $\tau_n^t$ is irreducible for $(n,t)\in \mathcal{D}_n$, and prove that $\tau_n^t$ is irreducible for $(n,t)\in \mathcal{D}_{n+1}$.
Let us define the set $\mathcal{I}_{n+1}=\mathcal{D}_{n+1}\setminus \mathcal{D}_n$ for $n\ge 1$.
We rewrite some elements in $\mathcal{I}_{n}\cup \mathcal{I}_{n+1}$ for simplicity as
\begin{align*}
&A_0=x_{n+1}, B_0=x_n, C_0=0, A_1=z_n, B_1=z_{n-1}, C_1=y_n,\\
& A_m=\tau_{\bar{m}}^{2m}, B_m=\tau_{\bar{m}-1}^{2m}, C_m=\tau_{\bar{m}-1}^{2m-1},
\end{align*}
where $1\le m\le n$ and $\bar{m}:=n+1-m$.
\begin{figure}
\centering
\includegraphics[width=6cm,bb=200 220 400 400]{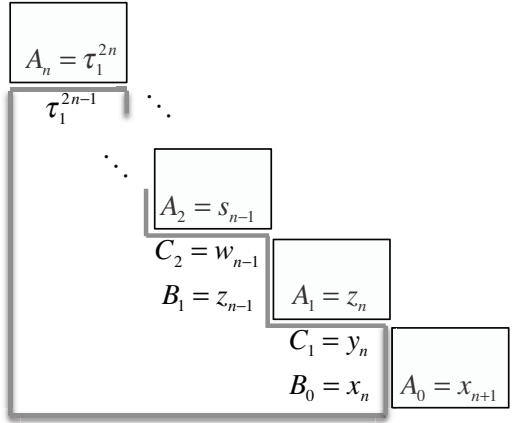}
\caption{Regions $\mathcal{D}_n$ and $\mathcal{I}_n$. The region $\mathcal{D}_n$ is enclosed by gray lines in the left hand side of the figure. The region $\mathcal{I}_{n+1}$ consists of collection of boxes over $\mathcal{D}_n$.}
\label{figure4}
\end{figure}
See figure \ref{figure4}. Then from equation \eqref{dtoda}, we have
\begin{equation}
A_m=\frac{1}{B_{m-1}}\left(B_m A_{m-1}+(C_{m})^2 \right)\;\; (m=1,2,\cdots, n). \label{Am}
\end{equation}
\begin{Lemma} \label{AmR}
We have $A_m \in R$.
\end{Lemma}
\textbf{Proof of lemma \ref{AmR}}\;\;
We show by induction.
Equation \eqref{Am} is equivalent to $a_3 c_3=a_2c_4+b_3^2$ with the notation in lemma \ref{c3a3}. Using lemma \ref{c3a3}, we have
\[
a_3 c_3=\frac{c_3}{c_2 d_3^2 e_4}\cdot P \in R,
\]
where $P$ is a polynomial term as in lemma \ref{c3a3}.
From the induction hypothesis that every pair of two terms is coprime in $\mathcal{D}_n$, the term $c_3=B_{m-1}$ in $\mathcal{D}_n$ is coprime with $c_2,d_3,e_4$. Therefore $P$ has to be divisible by $c_2 d_3^2 e_4$ in $R$. Thus $a_3\in R$.
\qed
Next we prove $A_m\in R_{irr}$. We use the following lemma \ref{AmRirred}, which can be proved inductively from \eqref{dtoda}:
\begin{Lemma} \label{AmRirred}
The term $A_m$ is a linear equation with respect to $x_{n+1}$.
When we write $A_m=\alpha_m x_{n+1} + \beta_m$, we have $\alpha_m, \beta_m \in R_{n,n}$ and $\alpha_m=B_m/x_n$.
\end{Lemma}
From the induction hypothesis, $B_m\in R_{irr}$. From the expression $A_m=\alpha_m x_{n+1}+\beta_m$ in lemma \ref{AmRirred}, we have from lemma \ref{fgirred} that `$A_m\in R_{irr}$ if $\beta_m$ does not have $B_m$ as one of its factors'. We prove this by contradiction.
Let us suppose that $\beta_m$ has the factor $B_m$. Then $B_m$ divides $A_m$. Equation \eqref{Am} is equivalent to
\[
A_m B_{m-1}=B_m A_{m-1}+ C_m^2,
\]
which indicates that $C_m^2$ has the factor $B_m$.
This contradicts the induction hypothesis that every pair of terms is coprime in $\mathcal{D}_n$. Therefore $A_m\in R_{irr}$.
The irreducibility of $\tau_{\bar{m}}^{2m+1}$ in $\mathcal{I}_{n+1}$ (the element $\tau_{\bar{m}}^{2m+1}$ is just above $A_m$ in $n$-$t$ plane) is proved in the same manner.
We also have that $A_m$ and $A_{m'}$ are coprime if $m\neq m'$, since the coefficients of $x_{n+1}$ in $A_m$ is $B_m/x_n$ (resp. in $A_{m'}$ is $B_{m'}/x_n$), and $B_m$ and $B_{m'}$ are coprime. Next since the elements $\tau_l^s$ in $\mathcal{D}_n$ does not contain $x_{n+1}$, we have that $\tau_l^s$ and $A_m$ are coprime. Thus we have proved the irreducibility and co-primeness for the elements in $\mathcal{I}_{n+1}$, and therefore in $\mathcal{D}_{n+1}$.
\qed
The next proposition states that the same result is true even for a specialized initial conditions $\tau_1^0=x_1=1$.
\begin{Proposition} \label{thmsemiinf1}
Every term of the discrete Toda equation \eqref{dtoda} is a Laurent polynomial of the initial variables:
\[
\tau_n^t\in\mathbb{Z}[x_2^{\pm}, x_3^{\pm}, \cdots, y_1^{\pm}, y_2^{\pm},\cdots],
\]
where $\tau_1^0=1$, $\tau_n^0=x_n$ $(n\ge 2)$ and $\tau_n^1=y_n$ $(n\ge 1)$.
Moreover, the term $\tau_n^t$ is an irreducible Laurent polynomial.
\end{Proposition}
\textbf{Proof}\;\;
The Laurent property
\[
\tau_n^t\in\mathbb{Z}[x_2^{\pm}, x_3^{\pm}, \cdots, y_1^{\pm}, y_2^{\pm},\cdots],
\]
is trivially obtained by substituting $\tau_1^0=1$ in theorem \ref{thmsemiinf}.
We now prove the irreducibility.
We consider the transformation $\tau_n^t=(x_1)^n \sigma_n^t$.
If $\tau_n^t$ satisfies the discrete Toda equation \eqref{dtoda}, 
new function $\sigma_n^t$ satisfies the same form:
\[
\sigma_n^{t+1} \sigma_n^{t-1} = \sigma_{n-1}^{t+1} \sigma_{n+1}^{t-1} + (\sigma_n^t)^2.
\]
The function $\sigma_n^t$ is obtained by developing equation \eqref{dtoda} from the initial values $\tau_1^0=1$, $\tau_n^0={x_n}/{(x_1)^n}\, (n\ge 2)$ and $\tau_n^1={y_n}/{(x_1)^n}\, (n\ge 1)$, and therefore satisfies
$\sigma_n^t=(x_1)^{-n} \tau_n^t$ for all $n,t$.
The irreducibility and co-primeness are preserved under the transformation
$\sigma_n^t=(x_1)^{-n} \tau_n^t$, since it is only a multiplication by a monomial.
\qed
\begin{Theorem} \label{todaIVsemiinfthm}
The solution $I_n^t$, $V_n^t$ of the discrete semi-infinite Toda equation
(\eqref{dtodaIV1} and \eqref{dtodaIV2} with $V_0^t=0 (t\ge 0)$) satisfies the following `co-prime' property:
Let us define the set $\mathcal{D}=\{I_n^t\}\cup\{V_n^t\}$. We denote by $D_n^t$ an element in $\mathcal{D}$ with scripts of time $t$ and position $n$.
Two distinct elements $D_n^t$ and $D_m^s$ in the set $\mathcal{D}$
do not have common factors other than monomials of the initial variables, on condition that $|n-m|\ge 3$ or $|t-s|\ge 2$.
\end{Theorem}
\textbf{Proof}\;\;
By equation \eqref{tauIV}, the correspondence of initial values between the bilinear and the nonlinear discrete Toda equation is as follows:
\begin{align*}
&I_1^0=y_1, V_1^0=\frac{x_2}{y_1}, I_2^0=\frac{y_2}{x_2 y_1}, V_2^0=\frac{x_3 y_1}{x_2 y_2}, I_3^0=\frac{x_2 y_3}{x_3 y_2}, V_3^0=\frac{x_4 y_2}{x_3 y_3},\\
&\cdots, V_{N-1}^0=\frac{x_N y_{N-2}}{x_{N-1} y_{N-1}}, I_N^0=\frac{x_{N-1} y_N}{x_N y_{N-1}},\cdots.
\end{align*}
The co-primeness in terms of $\tau_n^t$ proved in theorem \ref{thmsemiinf} and proposition \ref{thmsemiinf1} is transformed into the `co-primeness' of $I_n^t,V_n^t$ by equation \eqref{tauIV}.
For example, $I_n^t$ and $I_m^s$ share a certain $\tau_l^u$ in their numerators or denominators, if and only if
$|n-m|\le 1$ and $|t-s|\le 1$. Therefore $I_n^t$ and $I_m^s$ are coprime as rational functions (cf. definition \ref{rationalcoprime}), if and only if
\[
|n-m|\ge 2,\  \mbox{or},\  |t-s|\ge 2.
\]
In the same manner, $V_n^t$ and $V_m^s$ are coprime if and only if
\[
(m-n,s-t)\neq (\pm 1,0),(0,\pm 1),\pm(1,-1),\pm(2,-1).
\]
We also have that $I_n^t$ and $V_m^s$ are coprime if and only if
\[
(m-n,t-s)\neq (0,0),(\pm 1,0),(0,\pm 1),\pm(1,-1),(-1,-1),(-2,0),(-2,1).
\]
All these three conditions are satisfied if $|n-m|\ge 3$ or $|t-s|\ge 2$.
\qed
\subsection{Molecule boundary}
We impose the molecule boundary condition on the equation \eqref{dtoda} as
\[
\tau_{N+2}^t=0\ (t=0,1,2,\cdots),
\]
in addition to the conditions
\begin{equation} \label{moleculeboundary}
\tau_0^t=1\ (t\ge 0),\; \tau_n^0=x_n,\ \tau_n^1=y_n\ (1\le n\le N+1),
\end{equation}
where $N(\ge 1)$ is the system size of the discrete Toda molecule equation.
We study the irreducibility and co-primeness under this condition, and prove that statements very similar to those of theorems \ref{thmsemiinf} and \ref{todaIVsemiinfthm} hold.

\begin{Theorem} \label{thmmolecule}
Every term of the discrete molecule Toda equation \eqref{dtoda} with $\tau_{N+2}^t=0$ $(t=0,1,2,\cdots)$, and \eqref{moleculeboundary} is a Laurent polynomial of the initial variables:
\[
R:=\tau_n^t\in\mathbb{Z}[x_1^{\pm}, x_2^{\pm},\cdots, x_{N+1}^{\pm}, y_1^{\pm}, y_2^{\pm},\cdots, y_{N+1}^{\pm}].
\]
Moreover, the term $\tau_n^t$ is an irreducible Laurent polynomial.
\end{Theorem}
\textbf{Proof}\;\;
Define the set of irreducible Laurent polynomials as
\[
R_{irr}:=\{f\in R\ |\ f\ \mbox{is irreducible}\ \}.
\]
To ease notation, we use the same symbol $R$ of previous section for different rings.
First let us prove the Laurentness and then the irreducibility.
\begin{Lemma} \label{moleculelau}
We have $\tau_n^t\in R$.
\end{Lemma}
\textbf{Proof of lemma \ref{moleculelau}}
We already have the Laurent property for the discrete Toda equation with semi-infinite boundary condition in theorem \ref{thmsemiinf}.
The discrete Toda equation with molecule boundary condition is obtained by substituting $x_n=y_n=0$ for every $n\ge N+2$.
Let us take an arbitrary Laurent polynomial $f\in\mathbb{Z}[x_i^{\pm}, y_i^{\pm} ; 1\le i]$.
By substituting $x_n=y_n=0$ $(n\ge N+2)$ in $f$, we have either
\[
f|_{x_n=y_n=0\, (n\ge N+2)}\in\mathbb{Z}[x_i^{\pm}, y_i^{\pm} ;\, 1\le i\le N+1],
\]
or $f|_{x_n=y_n=0\, (n\ge N+2)}$ is not defined because of the zero denominator.
However, for $\tau_n^t$ ($1\le n\le N+1, \, t\ge 2$) here, we do not encounter zero in the denominators, since all the terms $\tau_n^t$ are well-defined by \eqref{dtoda} and the conditions $\tau_{N+2}^t=0$ and \eqref{moleculeboundary}.
\qed
\begin{Lemma} \label{moleculeirred}
We have $\tau_n^t\in R_{irr}$.
\end{Lemma}
\textbf{Proof of lemma \ref{moleculeirred}}\;\;
Let us rewrite $z_n:=\tau_n^2$, $u_n:=\tau_n^3$, $v_n:=\tau_n^4$,
$w_n:=\tau_n^5$, $s_n:=\tau_n^6$.
\paragraph{The case of $\boldsymbol{N=1}$:}
Terms $z_1$ and $u_1$ are the same as those for semi-infinite boundary condition, and therefore are irreducible.
Since we have $\tau_2^t=(y_2)^t /(x_2)^{t-1}$ for $t\ge 2$,
$\tau_2^t\in R_{irr}$ for all $t\ge 2$.
Therefore we only have to prove the irreducibility of $\tau_1^t$ $(t\ge 4)$, and their co-primeness with other terms.
The term $u_1$ is a function of $x_1,x_2,y_1,y_2$. If we substitute $x_i\to y_i$ and $y_i\to z_i$ in $u_1$, we obtain $v_1$:
\[
v_1=u_1 \big|_{x_i\to y_i, y_i\to z_i}.
\]
We use lemma \ref{locallemma} for $m=4$, $(p_1,p_2,p_3,p_4)=(y_1,y_2,z_1,z_2)$, $(q_1,q_2,q_3,q_4)=(x_1,x_2,y_1,y_2)$, and $f(y_1,y_2,z_1,z_2)=v_1$, to obtain
\[
v_1 = z_1^{r_1} z_2^{r_2} \cdot P,
\]
where $P\in R_{irr}$, $r_1,r_2\in\mathbb{Z}_{\ge 0}$. (cf. $f(x_1,x_2,y_1,y_2)=u_1$)
Now let us substitute $x_i=y_i=1$ $(i=1,2)$, to obtain $z_1=2,v_1=13$.
Therefore, $13$ should be divisible by $2^{r_1}$ in $\mathbb{Z}$. Thus we have $r_1=0$. Since $z_2=y_2^2/ x_2$ is a unit in $R$, we conclude that $v_1\in R_{irr}$.
We also have that $v_1$ is coprime with $z_1$ and $u_1$, because two irreducible Laurent polynomials with distinct degrees are coprime.
Next we prove that $w_1:=\tau_1^5 \in R_{irr}$.
We use lemma \ref{locallemma} for $m=4$, $(p_1,p_2,p_3,p_4)=(u_1,u_2,v_1,v_2)$, $(q_1,q_2,q_3,q_4)=(x_1,x_2,y_1,y_2)$, and $f(u_1,u_2,v_1,v_2)=w_1$, to obtain
\[
w_1 = u_1^{r_1} v_2^{r_2} u_2^{r_3} v_2^{r_4} \cdot P_2,
\]
where $P_2\in R_{irr}$, each $r_i\in\mathbb{Z}$.
Substituting $x_i=y_i=1$ $(i=1,2)$, we have
\[
34=5^{r_1} 13^{r_2} \cdot 1 \cdot 1\cdot p_2,
\]
where $p_2:=(P_2)|_{x_i=y_i=1}\in\mathbb{Z}$.
Therefore $r_1=r_2=0$. Together with the fact that $u_2,v_2$ are units in $R$, we have $w_1\in R_{irr}$.
In the same manner we have from lemma \ref{locallemma} that,
\[
s_1=v_1^{r_1} w_1^{r_2} v_2^{r_3} w_2^{r_4}\cdot P_3,
\]
where $P_3\in R_{irr}$, each $r_i\in\mathbb{Z}$ (To ease notation we used the same $r_i$ as before for different values).
Substituting $x_i=y_i=1$ $(i=1,2)$, we have
\[
89=13^{r_1} 34^{r_2}\cdot p_3,
\]
where $p_3\in\mathbb{Z}$. Thus $r_1=r_2=0$. Therefore $s_1\in R_{irr}$.
Co-primeness of $s_1$ with other elements can be proved in the same manner.
Finally we prove the case of $\tau_n^t$ $(t\ge 7)$.
We have the following three decompositions for $\tau_1^t$ $(t\ge 7)$:
\begin{eqnarray*}
\tau_1^t&=&z_1^{r_1} z_2^{r_2}\cdot Q_1\\
&=&u_1^{r_3} v_1^{r_4} u_2^{r_5} v_2^{r_6}\cdot Q_2\\
&=& w_1^{r_7} s_1^{r_8} w_2^{r_9} s_2^{r_{10}}\cdot Q_3,
\end{eqnarray*}
where $Q_i\in R_{irr}$, $r_i\in\mathbb{Z}$.
Since we have already proved that $z_1,u_1,v_1,w_1,s_1$ are irreducible and coprime with each other, we have $r_1=r_3=r_4=r_7=r_8=0$.
Note that $\tau_n^2$ is a unit in $R$. Thus we have $\tau_1^t\in R_{irr}$ for $t\ge 7$.
\paragraph{The case of $\boldsymbol{N=2}$:}
The proof is very similar to that of $N=1$ case.
Since $\tau_3^t$ is a unit in $R$ for every $t$, we prove the irreducibility of $\tau_1^t$ and $\tau_2^t$. Co-primeness between two terms are proved by investigating the degrees of the terms.
We use lemma \ref{locallemma} repeatedly and then substitute $x_i=y_i=1$ $(i=1,2)$.
Note that
\begin{align*}
&(x_1,y_1,z_1,u_1,v_1,w_1,s_1)=(1,1,2,5,14,42,131),\\
&(x_2,y_2,z_2,u_2,v_2,w_2,s_2)=(1,1,3,14,70,353,1782).
\end{align*}
\paragraph{The case of $\boldsymbol{N\ge 3}$:}
First, we prove the irreducibility of $\tau_n^t$ for $2\le n\le N+1$ and $t\le N$.
For $t\le 6$, we have only to prove the irreducibility of the four terms $v_N,w_N,s_N,s_{N-1}$, since other terms $\tau_n^t$ with $2\le n\le N$ are the same as in the case of semi-infinite boundary condition, and $\tau_{N+1}^t$ is a unit in $R$ for every $t\ge 0$.
Let us prove the irreducibility of these four terms individually.
\subparagraph{The case of $\boldsymbol{v_N}$:}
First we prove that $v_N\in R_{irr}$.
With some calculation we have the following expression for $v_N$:
\begin{equation} \label{vN}
v_N=\frac{1}{z_N}\left( \frac{v_{N-1}}{x_{N+1}}+\frac{(u_{N-1})^2}{(y_N)^2}\right) (y_{N+1})^2+\frac{2 u_{N-1} z_N}{(y_N)^2}y_{N+1} +\frac{(z_N)^3}{(y_N)^2}.
\end{equation}
Let us rewrite the coefficients of $(y_{N+1})^2$ as $G_N$ and obtain the recurrence relation for $G_N$ as follows:
\begin{align*}
G_N&=\frac{1}{z_N}\left( \frac{v_{N-1}}{x_{N+1}}+\frac{(u_{N-1})^2}{(y_N)^2}\right)=\frac{x_N}{x_{N+1} z_{N-1}} \left( \frac{v_{N-2}}{x_{N}}+\frac{(u_{N-1})^2}{(y_N)^2}\right)\\
&=\frac{x_N}{x_{N+1} z_{N-1}}\left( \frac{v_{N-2}}{x_{N}}+\frac{(u_{N-2})^2}{(y_{N-1})^2}+\frac{(u_{N-1})^2}{(y_N)^2}- \frac{(u_{N-2})^2}{(y_{N-1})^2}\right)\\
&=\frac{x_N}{x_{N+1} z_{N-1}}\left( \frac{v_{N-2}}{x_{N}}+\frac{(u_{N-2})^2}{(y_{N-1})^2}+ \frac{(z_{N-1})^2}{y_{N-1} y_N}\left( \frac{u_{N-1}}{y_N}+\frac{u_{N-2}}{y_{N-1}} \right) \right)\\
&=\frac{x_N}{x_{N+1}}G_{N-1}+\frac{z_{N-1} x_N}{x_{N+1} y_{N-1} y_N}\left( \frac{u_{N-1}}{y_N}+\frac{u_{N-2}}{y_{N-1}}\right).
\end{align*}
Here we have used in the first equality the following relations obtained from \eqref{dtoda}:
\[
v_{N-1}=\frac{(u_{N-1})^2+z_N v_{N-2}}{z_{N-1}},\; z_N=\frac{(y_N)^2+z_{N-1} x_{N+1}}{x_N}.
\]
By using this recurrence relation we obtain
\begin{equation} \label{GNrec}
x_{N+1} G_N=x_2 G_1+\sum_{k=1}^{N-1} \frac{z_k x_{k+1}}{y_k y_{k+1}}\left( \frac{u_k}{y_{k+1}}+\frac{u_{k-1}}{y_k} \right),
\end{equation}
where $G_1=\frac{x_1}{x_2(y_1)^2}$, $u_0=1$. Since the right hand side of
\eqref{GNrec} does not depend on $x_{N+1}$, we can express $G_N$ as
\begin{equation} \label{GNGamma}
G_N=\frac{\Gamma_N}{x_{N+1}},
\end{equation}
where $\Gamma_N$ does not contain $x_{N+1}$.
We have that $\Gamma_N$ does not have the factor $z_N$, since $z_N$ is linear with respect to $x_{N+1}$, whose constant term $\frac{y_N^2}{x_N}$ is nonzero.
Therefore if we suppose that $v_N$ is not irreducible, only the following type of decomposition is possible:
\[
v_N=(a y_{N+1}+ b)(cy_{N+1}+d),
\]
where $a,b,c,d\in R_{N+1,N}$, because the decomposition of the type $v_N=z_N\cdot P$ ($P\in R$) is not possible from equation \eqref{GNGamma}.
We prove that $b/z_n$ and $d/(z_n)^2$ are both units in $R$.
Since $a,c\in R_{N+1,N}$, the terms $a,c$ do not have a factor $z_N$.
From $bd=z_N^3/y_N^2$, and from the irreducibility of $z_N$, we can decompose $(bd)$ as $b=\beta z_N^k$, $d=\beta' z_N^{3-k}$, where $k\in\mathbb{Z}$ and $\beta,\beta'$ are units in $R$. (Note that $y_N$ is also a unit in $R$.) Since $ad+bc$ has a factor $z_N$ as $(z_N)^1$, we have $\min [k,3-k]=1$. Therefore $k=1$ or $k=2$.
We can choose $k=1$ without losing generality and write down $b,d$ as
\[
b=\gamma z_N,\;\; d=\frac{(z_N)^2}{\gamma (y_N)^2},
\]
where $\gamma$ is a unit in $R$. This expression, together with $ac=\Gamma_N /x_{N+1}$ from \eqref{GNGamma}, indicates that the
coefficient of $y_{N+1}$ in equation \eqref{vN} satisfies
\[
\frac{(ac+bd)}{z_N}=\left(a\frac{z_N}{\gamma (y_N)^2}+c\gamma\right)=\frac{2 u_{N-1}}{(y_N)^2}.
\]
Since the right hand side does not depend on $x_{N+1}$, and therefore on $z_N$, while the middle term depends on $z_N$, we reach a contradiction. Therefore we conclude that $v_N$ is irreducible.
\subparagraph{The case of $\boldsymbol{w_N}$:}
By using lemma \ref{locallemma} we obtain the following two types of decomposition for $w_N$:
\begin{align*}
w_N&=z_1^{r_1}\cdots z_{N+1}^{r_{N+1}}\cdot P\\
&=u_1^{s_1}\cdots u_{N+1}^{s_{N+1}}\cdot v_1^{q_1}\cdots v_{N+1}^{q_{N+1}}\cdot Q,
\end{align*}
where $P,Q\in R_{irr}$ and $r_i,s_i,q_i\in\mathbb{Z}$.
Since each $z_i,u_i,v_i$ are irreducible and coprime with each other,
the only possible decompositions are one of the two types:
\begin{equation}
w_N=\delta z_i v_j,\ w_N=\delta z_i u_j, \label{wdecomp}
\end{equation}
where $\delta$ is a unit in $R$.
We prove that none of the two decomposition is possible by investigating the degrees of the terms in $y_{N+1}$.
Let us denote deg $f$ as the degree of $f$ as a polynomial of $y_{N+1}$.
We have deg $w_N$=3, deg $v_{N+1}=4$, deg $v_N=2$, deg $u_{N+1}=3$, deg $u_N=1$, deg $z_{N+1}=2$, deg $v_i=$ deg $u_i=0\, (i\le N-1)$, deg $z_i=0\, (i\le N)$.
For the degrees to be equal in both sides of the equation \eqref{wdecomp}, we have the following two possibilities:
\[
w_N=\delta u_N z_{N+1},
\]
or
\[
w_N=\delta u_{N+1} z_i, (i\le N).
\]
Note that the unit $\delta$ does not depend on $y_{N+1}$, since we easily verify that the constant term of $w_N$ as a polynomial of $y_{N+1}$ is nonzero.
However, two terms $z_{N+1}=y_{N+1}^2/x_{N+1}$ and $u_{N+1}=y_{N+1}^3/x_{N+1}^2$ are both monomials of $y_{N+1}$. These facts contradict the nonzero constant term of $w_N$.
Therefore none of the two decomposition is possible,  and we have proved that $w_N\in R_{irr}$. Co-primeness with other terms is also proved by investigating the degrees of the terms.
\subparagraph{The case of $s_{N-1}, s_N$:}
Other two terms $s_N,s_{N-1}$ are proved to be irreducible in similar discussions.
Lemma \ref{locallemma} gives the following two types of decomposition for $s_{N-1}$:
\begin{align*}
s_{N-1}&=u_1^{r_1}\cdots u_{N+1}^{r_{N+1}}\cdot P\\
&=v_1^{s_1}\cdots v_{N+1}^{s_{N+1}}\cdot w_1^{q_1}\cdots w_{N+1}^{q_{N+1}}\cdot Q,
\end{align*}
where $P,Q\in R_{irr}$ and $r_i,s_i,q_i\in\mathbb{Z}$.
Since each $u_i,v_i,w_i$ are irreducible and coprime with each other,
the only possible decompositions are one of the two types:
\begin{equation}
s_{N-1}=\delta u_i v_j,\ s_{N-1}=\delta u_i w_j, \label{sdecomp1}
\end{equation}
where $\delta$ is a unit in $R$.
By investigating the degrees of these terms as polynomials of $y_{N+1}$, only the following two cases are possible:
\begin{eqnarray}
s_{N-1}&=&\delta u_i v_N\, (i\le N-1), \label{sdecomp2}\\
\mbox{or} \notag \\
s_{N-1}&=&\delta u_N w_{N-1}. \label{sdecomp3}
\end{eqnarray}
From \eqref{dtoda} we have $s_{N-1} v_{N-1}=w_{N-1}^2+s_{N-2}v_N$.
The first equation \eqref{sdecomp2} gives
\[
(\delta u_i v_{N-1}-s_{N-2})v_N=w_{N-1}^2,
\]
which is a contradiction because of the irreducibility of $w_{N-1}$ proved in the previous paragraph.
The second one \eqref{sdecomp3} is also a contradiction, since it gives
\[
(\delta u_N v_{N-1}- w_{N-1})w_{N-1}=s_{N-2} v_N,
\]
and every pair of terms here is coprime. Therefore both decompositions in \eqref{sdecomp1} are impossible, and thus $s_{N-1}\in R_{irr}$.
As for the term $s_N$, by the same investigations, we obtain the three possible factorizations:
\[
s_N=\delta u_N w_N,\ \mbox{or}\ s_N =\delta u_{N+1} w_{N-1},\ \mbox{or}\ s_N=\delta u_i v_{N+1}\, (i\le N-1),
\]
none of which turns out to be possible. By substituting $s_N v_N=w_N^2+s_{N-1} v_{N+1}$ in the first equation $s_N=\delta u_N w_N$, we obtain
\[
(\delta u_N v_N-w_N)w_N=s_{N-1} v_{N+1},
\]
which is impossible from the irreducibility and co-primeness of the terms. Note that we used here the irreducibility of $s_{N-1}$, which has just been proved.
The latter two equations are impossible because the relations $u_{N+1}=y_{N+1}^3/x_{N+1}^2$ and $v_{N+1}=y_{N+1}^4/x_{N+1}^3$ contradict the fact that $s_N$ has a nonzero constant term as a polynomial of $y_{N+1}$.
Thus $s_N\in R_{irr}$.
We have proved the irreducibility of $\tau_n^t$ for $t\le 6$.

Finally we prove the case for $t\ge 7$.
We have the following three types of decompositions of $\tau_n^t$ for $t\ge 7$:
\begin{align*}
\tau_n^t&= z_1^{r_{1,1}}\cdots z_{N+1}^{r_{1,N+1}} P_1\\
&=u_1^{r_{2,1}}\cdots u_{N+1}^{r_{2,N+1}} v_1^{r_{3,1}}\cdots v_{N+1}^{r_{3,N+1}}P_2\\
&=w_1^{r_{4,1}}\cdots w_{N+1}^{r_{4,N+1}} s_1^{r_{5,1}}\cdots s_{N+1}^{r_{5,N+1}}P_3.
\end{align*}
Here, each $r_{i,j}\in\mathbb{Z}$ and $P_i\in R_{irr}$.
Since any pair from $\{u_i,v_i,w_i,s_i\}$ is coprime, this decomposition is only possible when
$r_{i,j}=0$ for all $i,j$. Therefore $\tau_n^t\in R_{irr}$.
Thus theorem \ref{thmmolecule} is proved.
\qed
We have the following proposition for a specialized initial condition $\tau_1^0=x_1=1$:
\begin{Proposition}
Every term of the discrete Toda equation \eqref{dtoda} is a Laurent polynomial of the initial variables:
\[
\tau_n^t\in\mathbb{Z}[x_2^{\pm}, x_3^{\pm}, \cdots, x_{N+1}^{\pm}, y_1^{\pm}, y_2^{\pm},\cdots, y_{N+1}^{\pm}],
\]
where $\tau_n^0=x_n$ $(2\le n\le N+1)$ and $\tau_n^1=y_n$ $(1 \le n\le N+1)$.
Moreover, the term $\tau_n^t$ is an irreducible Laurent polynomial.
\end{Proposition}
\textbf{Proof}\;\;
The proof is just the same as that of proposition \ref{thmsemiinf1}.\qed

\begin{Theorem}
The solution $I_n^t$, $V_n^t$ of the discrete molecule Toda equation
(\eqref{dtodaIV1} and \eqref{dtodaIV2} with $V_0^t=0$,$V_{N+1}^t=0$ $(t\ge 0)$) satisfies the following `co-prime' property:
Let us define the set $\mathcal{D}=\{I_n^t\}\cup\{V_n^t\}$.
Two distinct elements $D_n^t$ and $D_m^s$ in the set $\mathcal{D}$
do not have common factors other than monomials of the initial variables, on condition that $|n-m|\ge 3$ or $|t-s|\ge 2$.
\end{Theorem}
\textbf{Proof}\;\; The proof is just the same as in theorem \ref{todaIVsemiinfthm}.
\qed
%%%
%%%
%%%
%%%
%%%
\subsection{Periodic boundary}
We can obtain a co-primeness property similar to those in previous two sections for periodic discrete Toda equation, with more elaborated discussion.
Here the periodic boundary condition is imposed on the system \eqref{dtodaIV1} and \eqref{dtodaIV2} as follows:
\begin{equation}
I_{n+N}^t=I_n^t,\;\;V_{n+N}^t=V_n^t \label{IVperiod}
\end{equation}
for every $t$ and $n$, where $N$ is a positive integer which determines the system size.
\begin{Lemma} \label{lemmaperiod}
Let us suppose that $\prod_{i=1}^N V_i^t \neq \prod_{i=1}^N I_i^t$.
The time evolution of the periodic discrete Toda system \eqref{dtodaIV1}, \eqref{dtodaIV2} with \eqref{IVperiod} is determined by
\[
I_n^{t+1}=V_n^t+I_n^t Y_n^t,\ V_n^{t+1}=\frac{I_{n+1}^t V_n^t}{V_n^t+I_n^t Y_n^t},
\]
where
\[
Y_n^t=\frac{\left(1-\frac{\prod_{i=1}^N V_i^t}{\prod_{i=1}^N I_i^t}\right)}{1+\frac{V_{n-1}^t}{I_{n-1}^t}+\frac{V_{n-1}^t V_{n-2}^t}{I_{n-1}^t I_{n-2}^t}+\cdots +\frac{V_{n-1}^t V_{n-2}^t \cdots V_{n+1}^t}{I_{n-1}^t I_{n-2}^t \cdots I_{n+1}^t}}.
\]
Aside from the trivial solution $I_n^{t+1}=V_n^t$, $V_n^{t+1}=I_{n+1}^t$, this is the only solution for a fixed set of initial data $\{V_i^0, I_i^0\}_{i=1}^N$.
\end{Lemma}
We have to take care that the function $\tau_n^t$ does not necessarily satisfy the periodic condition $\tau_{n+N}^t=\tau_n^t$.
The reason is as follows. If we were to impose $\tau_{n+N}^t=\tau_n^t$, then we have
\[
\prod_{i=1}^N I_i^t=\prod_{i=1}^N \frac{\tau_{i-1}^t \tau_i^{t+1}}{\tau_i^t \tau_{i-1}^{t+1}}=\frac{\tau_N^{t+1} \tau_0^t}{\tau_0^{t+1}\tau_N^t}=1.
\]
In the same manner, we have $\prod_{i=1}^N V_i^t=1$.
However, the discrete Toda equation \eqref{dtodaIV1} and \eqref{dtodaIV2} cannot be determined under the condition \eqref{IVperiod} in the case of $\prod_{i=1}^N V_i^t=\prod_{i=1}^N I_i^t$ from lemma \ref{lemmaperiod}.
In fact it is reasonable to take the boundary condition as follows:
\begin{equation}
\tau_{n+N}^t=K \lambda^t \mu^n \tau_n^t,\; \tau_0^0=\tau_1^0=\tau_0^1=1 \label{period}
\end{equation}
where
\begin{eqnarray}
K&=&\prod_{i=1}^N (V_i^0 I_i^0)^{N-i}, \label{K}\\
\mu&=&\prod_{i=1}^N V_i^0 I_i^0, \label{mu}\\
\lambda&=& \prod_{i=1}^N I_i^0. \label{lambda}
\end{eqnarray}
This condition is obtained as follows.
First we assume that the function $\tau_n^t$ obeys the rule \eqref{period}, and then show that the constants $K$, $\lambda$ and $\mu$ can be determined uniquely in compatible with the evolution of the systems.
We obtain $\tau_n^0$ and  $\tau_n^1$ $(n\ge 1)$ inductively from \eqref{tauIV} as follows:
\begin{eqnarray}
\tau_1^1&=&I_1^0,\; \tau_2^0=V_1^0 I_1^0,\; \tau_2^1=I_1^0 I_2^0 \tau_2^0,\cdots\\
\tau_n^0&=&\prod_{i=1}^{n-1} \left( V_i^0 I_i^0 \right)^{n-i},\;\;(n\ge 2), \label{taun0} \\
\tau_n^1&=&\left(\prod_{i=1}^n I_i^0 \right) \tau_n^0,\;\;(n\ge 1). \label{taun1}
\end{eqnarray}
Using \eqref{taun0} for $n=N$, we obtain the value of $K=K\tau_0^0=\tau_N^0$ as in \eqref{K}.
Since $\tau_{N+1}^0=K\mu \tau_1^0=K\mu$, we have from \eqref{taun0} the equality \eqref{mu}.
From equation \eqref{taun1}, $\tau_N^1=K\lambda$ and $\tau_N^0=K$, we obtain \eqref{lambda}.

\begin{Proposition}
The function $\tau_n^t$ defined by
\begin{equation} \label{periodtau3}
\tau_n^{t+1}=\frac{\tau_{n+1}^{t-1}}{\lambda^2/\mu -1} \sum_{k=1}^N \frac{(\tau_{n+k}^t)^2}{\tau_{n+k}^{t-1} \tau_{n+k+1}^{t-1}}
\end{equation}
satisfies the bilinear form of the discrete Toda equation \eqref{dtoda} and also the periodic boundary condition \eqref{period}.
What is more, functions $I_n^t$ and $V_n^t$ obtained by the relation \eqref{tauIV} satisfies the discrete Toda equations \eqref{dtodaIV1} and \eqref{dtodaIV2}.
\end{Proposition}
\textbf{Proof}\;\;
We easily show by induction that $\tau_n^t$ defined by \eqref{periodtau3} satisfies the relation $\tau_{n+N}^t=K \lambda^t \mu^n \tau_n^t$ in \eqref{period}.
Next we show that \eqref{periodtau3} satisfies the discrete Toda equation \eqref{dtoda}:
\begin{align*}
&\tau_n^{t+1} \tau_n^{t-1}-\{\tau_{n-1}^{t+1} \tau_{n+1}^{t-1} +(\tau_n^t)^2 \}=\\
&\left( \frac{\tau_{n+1}^{t-1} }{ \lambda^2 / \mu-1} \sum_{k=1}^N \frac{(\tau_{k+n}^t)^2}{\tau_{k+n}^{t-1} \tau_{k+n+1}^{t-1}} \right) \tau_n^{t-1}- \left\{ \left( \frac{\tau_{n}^{t-1} }{ \lambda^2 / \mu-1} \sum_{k=1}^N \frac{(\tau_{k+n-1}^t)^2}{\tau_{k+n-1}^{t-1} \tau_{k+n}^{t-1}} \right) \tau_{n+1}^{t-1}+(\tau_n^t)^2\right\}=\\
& \frac{\tau_n^{t-1} \tau_{n+1}^{t-1} }{\lambda^2/\mu -1}\left\{ \frac{(\tau_{n+N}^t)^2}{\tau_{n+N}^{t-1}\tau_{n+N+1}^{t-1}}-\frac{(\tau_n^t)^2}{\tau_n^{t-1} \tau_{n+1}^{t-1}} \right\}-(\tau_n^t)^2=\\
& \frac{\tau_n^{t-1} \tau_{n+1}^{t-1} }{\lambda^2/\mu -1}\left\{ \frac{(K\lambda^t \mu^n)^2}{K\lambda^{t-1}\mu^n\cdot K\lambda^{t-1} \mu^{n+1}}-1 \right\}\frac{(\tau_n^t)^2}{\tau_n^{t-1} \tau_{n+1}^{t-1}}-(\tau_n^t)^2=0.
\end{align*}
\qed
Note that the equality \eqref{periodtau3} can be re-written as
\begin{equation} \label{periodtau4}
\tau_n^{t+1}=\frac{\tau_{n+1}^{t-1}}{\lambda^2/\mu -1} \left[ \frac{\lambda^2}{\mu}\sum_{j=0}^{n} \frac{(\tau_j^t)^2}{\tau_j^{t-1} \tau_{j+1}^{t-1} } +\sum_{j=n+1}^{N-1} \frac{(\tau_j^t)^2}{\tau_j^{t-1} \tau_{j+1}^{t-1}} \right],
\end{equation}
using the boundary condition \eqref{period}.

What we are going to prove is that ``The function $\tau_n^t$ is an irreducible Laurent polynomial of the initial variables ($\tau_n^0,\tau_n^1$; $0\le n \le N-1$) \textbf{and} a power of $(\lambda^2/\mu -1)$''.
To eliminate a power of $\lambda^2/\mu -1$, we change the variables.
\begin{Lemma}
The new variable $\tilde{\tau}_n^t$ defined by the transformation
\begin{equation} \label{tautilda}
\tau_n^t=\left(\frac{\lambda^2}{\mu}-1\right)^{-t(t-1)/2}\tilde{\tau}_n^t
\end{equation}
satisfies the following equation:
\begin{equation} \label{dtoda5.5}
\tilde{\tau}_n^{t+1} \tilde{\tau}_n^{t-1} = \tilde{\tau}_{n-1}^{t+1} \tilde{\tau}_{n+1}^{t-1} + \left(1-\frac{\lambda^2}{\mu}\right)(\tilde{\tau}_n^t)^2.
\end{equation}
\end{Lemma}
Note that the initial conditions are unchanged: $\tilde{\tau}_n^0=\tau_n^0$, $\tilde{\tau}_n^1=\tau_n^1$, since $t(t-1)/2=0$ for $t=0,1$. Also note that $\tau_N^0=K$, $\tau_N^1=K \lambda$ from the boundary condition \eqref{period}.
We are going to prove the Laurent property of this function $\tilde{\tau}_n^t$.
\begin{Theorem} \label{periodthm}
Let $k$ be an arbitrary natural number.

(A) The general term $\tilde{\tau}_n^t$ $(1\le n \le N$, $0\le t\le k)$ of the equation \eqref{dtoda5.5}
is in the following ring of Laurent polynomial
\[
\tilde{\tau}_n^t\in R:=\mathbb{Z}\left[(\tilde{\tau}_n^0)^{\pm}; 2\le n\le N-1,(\tilde{\tau}_n^1)^{\pm}; 1\le n \le N-1, K^{\pm},\lambda^{\pm},\mu^{\pm} \right].
\]

(B) Moreover $\tilde{\tau}_n^t$ is irreducible for any $n,t$ $(1\le n\le N$, $0\le t\le k)$, and two distinct terms $\tilde{\tau}_n^t$ and $\tilde{\tau}_m^s$ with $(n,t)\neq (m,s)$ are co-prime in this ring $R$.
\end{Theorem}
\textbf{Proof}\;\;We prove theorem  \ref{periodthm} by induction using the following propositions and lemmas. We rewrite $\tilde{\tau}$ as $\tau$ to simplify the notation. We also use $a_n:=\tau_n^0$, $b_n:=\tau_n^1$, $c_n:=\tau_n^2$, $\cdots$. We have
\[
R=\mathbb{Z}[(a_n)^{\pm},(b_n)^{\pm};\; 1\le n\le N].
\]
\begin{Proposition} \label{AB}
If both of the statements (A) and (B) are satisfied for a fixed $k\ge 1$, then
the statement (A) is true for $k+2$.
\end{Proposition}
\textbf{Proof of proposition \ref{AB}}\;\;
Note that (A) is trivial for $k=0,1,2,3$, and (B) is for $k=0,1$.
First let us prove (B) for $k=2$, and then prove (A) for $k=4$ using ``(B) for $k=2$'', and then prove the statement for general $k$.
\paragraph{Proof of (B) for $\boldsymbol{k=2}$}
By making the transformation \eqref{tautilda},
the factor $\lambda^2/\mu-1$ is eliminated:
\begin{equation} \label{periodtau3kai}
\tau_n^{t+1}=\tau_{n+1}^{t-1} \left[ \frac{\lambda^2}{\mu}\sum_{j=0}^n \frac{(\tau_j^t)^2}{\tau_j^{t-1} \tau_{j+1}^{t-1}} + \sum_{j=n+1}^{N-1} \frac{(\tau_j^t)^2}{\tau_j^{t-1} \tau_{j+1}^{t-1}} \right].
\end{equation}
Next, by substituting $t=1$ in equation \eqref{periodtau4}, we have
\begin{equation} \label{cn}
c_n=a_{n+1}\left[ \frac{\lambda^2}{\mu}\sum_{j=0}^n \frac{(b_j)^2}{a_j a_{j+1}} + \sum_{j=n+1}^{N-1} \frac{(b_j)^2}{a_j a_{j+1}} \right].
\end{equation}
We have that the term $c_n$ is irreducible in the ring $R$. 
We prove this by contradiction. If $c_n$ is reducible, it has to be factored as $(b_1+\alpha)(b_1+\beta)$ as a quadratic function of $b_1$, where $\alpha, \beta$ are expressed by $a_j$ and $b_k\, (k\neq 1)$. Since $c_n$ does not have a term of $(b_1)^1$, we have $\alpha=-\beta$. Therefore the constant term of $c_n$ in terms of $b_1$ is negative ($-\alpha^2<0$), which contradicts the fact that every coefficient is non-negative in $c_n$. Therefore $c_n=\tau_n^2$ is irreducible.
\paragraph{Proof of (A) for $\boldsymbol{k=4}$}
By shifting the superscripts to $t=3$ for equation \eqref{cn}, we obtain the following equality:
\begin{equation} \label{en}
e_n=c_{n+1}\left[ \frac{\lambda^2}{\mu}\sum_{j=0}^n \frac{(d_j)^2}{c_j c_{j+1}} + \sum_{j=n+1}^{N-1} \frac{(d_j)^2}{c_j c_{j+1}} \right].
\end{equation}
We prove that $e_n\in R$ for all $0\le n\le N-1$. We only have to prove that $e_0\in R$, since the subscripts are cyclic modulo $N$.
Reducing to a common denominator of \eqref{en}, we have
\begin{align}
&c_0 c_2 \cdots c_{N-1} e_0 = \notag \\
&\frac{\lambda^2}{\mu} (d_0)^2 c_2 c_3 \cdots c_{N-1}+c_0 (d_1)^2 c_3 \cdots c_{N-1}+c_0c_1(d_2)^2c_4\cdots c_{N-1}+ \notag \\
&\cdots +c_0c_1c_2 \cdots c_{N-3} (d_{N-2})^2+\frac{c_0}{c_N} c_1c_2\cdots c_{N-2} (d_{N-1})^2. \label{c0e0}
\end{align}
Since $\frac{c_0}{c_N}=\frac{1}{K\lambda^2}$ from the boundary condition \eqref{period}, the right hand side is a Laurent polynomial (i.e., $\in R$). Thus $c_0c_2c_3\cdots c_{N-1} e_0\in R$.

Next let us prove that $c_2c_3\cdots c_{N-1}e_0\in R$.
Let us pick up all the terms which do not contain $c_0$ from the right hand side of \eqref{c0e0} and define it as $E$:
\begin{equation}
E:=\left(\frac{\lambda^2}{\mu}(d_0)^2 c_{N-1} + \frac{1}{K\lambda^2}c_1 (d_{N-1})^2\right)\cdot c_2c_3\cdots c_{N-2}. 
\end{equation}
We prove that $E$ itself has a factor $c_0$.
From equation \eqref{periodtau3kai} with $(n,t)=(0,2)$ and $(n,t)=(N-1,2)$, we have
\[
\frac{d_0}{b_1}=\frac{\lambda^2}{\mu}\frac{(c_0)^2}{b_0 b_1}+\sum_{j=1}^{N-1} \frac{(c_j)^2}{b_j b_{j+1}},\;\; \frac{d_{N-1}}{b_N}=\frac{\lambda^2}{\mu} \sum_{j=0}^{N-1} \frac{(c_j)^2}{b_j b_{j+1}}.
\]
By substituting these equations to $E$, we obtain
\begin{equation} \label{E6}
E/(c_2c_3 \cdots c_{N-2})=(c_0)^2\cdot P +\frac{\lambda^2}{\mu}\left[(b_1)^2 c_{N-1}+ \frac{1}{K\lambda^2} (b_N)^2 c_1\right]\left( \sum_{j=1}^{N-1} \frac{(c_j)^2}{b_j b_{j+1}} \right)^2,
\end{equation}
where $P\in R$.
From the evolution of the equation \eqref{dtoda5.5} (note that we have omitted $\tilde{\ }$ here), we have
\[
c_1=\left\{a_2 c_0 +\left(1-\frac{\lambda^2}{\mu}\right)(b_1)^2\right\}\frac{1}{a_1},\;\;
c_{N-1}=\left\{a_N c_N - \left(1-\frac{\lambda^2}{\mu}\right)(b_N)^2\right\}\frac{1}{K \mu a_1},
\]
where we have used $a_{N+1}=K \mu a_1$.
Therefore we obtain
\begin{equation} \label{E6sup}
(b_1)^2 c_{N-1}+ \frac{1}{K\mu} (b_N)^2 c_1=\frac{K \lambda^2 (a_0 b_1^2 + a_2 b_0^2)}{\mu a_1}\cdot c_0.
\end{equation}
Using the equations \eqref{E6} and \eqref{E6sup}, we have proved that
$E/c_0\in R$. Therefore we have
\[
c_2c_3\cdots c_{N-1}e_0=(c_0 c_2c_3\cdots c_{N-1}e_0)/c_0= (E+\mathcal{O}(c_0))/c_0\in R.
\]
%%%
By a cyclic permutation, we have $(c_0 c_2 \cdots c_{N-1} e_0)/c_j \in R$ for all $0\le j\le N-1$.
From these results for all $j$, and from the fact that $c_j$ are irreducible for all $j$ (which has been proved as [(B) for $k=2$] in the previous paragraph), we have
\[
e_0=\frac{c_0 c_2 \cdots c_{N-1} e_0}{c_0c_2\cdots c_{N-1}}\in R.
\]
\paragraph{Proof of proposition \ref{AB} for general $\boldsymbol{k}$}
By shifting the time variable $t$ from $t=2$ to $t=k+1$ in the previous paragraph, we can prove that
\begin{equation} \label{tauk+2}
\left( \tau_0^k \tau_2^k\cdots \tau_{N-1}^k \right)\tau_0^{k+2}\in R.
\end{equation}
We also obtain
\[
\tau_0^{k+2}=\frac{L}{M},
\]
where $L,M\in R$ and $M$ is a monomial in $\{\tau_j^{k-1}, \tau_j^{k-2}\}_{j=1}^N$,
by shifting the time variable $t$ from $t=2$ to $t=k+1$ in equations from \eqref{en} through \eqref{E6sup} ($e_n\to \tau_n^{k+2}$, $b_n\to \tau_n^{k-1}$, $a_n\to \tau_n^{k-2}$).
Let us suppose that (B) is true for $k$, and define $P:=\tau_0^k \tau_2^k\cdots \tau_{N-1}^k$.
Then, by the irreducibility of each element, we have that $P$ and $M$ are coprime in $R$. On the other hand, we have $P\dfrac{L}{M}\in R$ from \eqref{tauk+2}, which indicates that $M$ must divide $L$ in the ring of Laurent polynomial $R$. Therefore $L/M \in R$.
We have proved (A) for $k\to k+2$, i.e., $\tau_0^{k+2}$ is a Laurent polynomial in $\{a_n,b_n\}_{n=0}^{N-1}$.
\qed
Using the proposition \ref{AB} repeatedly, the theorem \ref{periodthm} is derived by proving the following proposition:
\begin{Proposition} \label{kiyakulemmaperiod}
Let us assume that (A) is true for all $k\ge 1$.
Then (B) is true for all $k\ge 1$.
\end{Proposition}
\textbf{Proof of proposition \ref{kiyakulemmaperiod}}\;\;
We prove the irreducibility of $\tau_n^k$ for $k\ge 3$, since the case of $k=0,1$ is trivial, and the case of $k=2$ is already proved in the proof of proposition \ref{AB}.
\paragraph{The case  of $\boldsymbol{k=3}$:}
Let us apply lemma \ref{locallemma} in the case of $m=2N$,
\begin{eqnarray*}
\{p_1, \cdots,p_m\}&=&\{b_1,\cdots, b_N,c_1,\cdots,c_N\},\\
\{q_1, \cdots, q_m\}&=&\{a_1,\cdots, a_N, b_1,\cdots, b_N\}.
\end{eqnarray*}
From (A), each $d_j$ is irreducible in $\mathbb{Z}[\{b_i^{\pm}\},\{c_i^{\pm}\};1\le i\le N]$.
Therefore $d_j$ is decomposed as
\[
d_j=c_0^{r_0} c_1^{r_1} \cdots c_{N-1}^{r_{N-1}} \cdot G,
\]
where $G$ is an irreducible Laurent polynomial in $\mathbb{Z}[\{a_i^{\pm}\},\{b_i^{\pm}\}]$, and each $r_j\in\mathbb{Z}$. Since $c_j$ is irreducible, and $d_j$ is a Laurent polynomial, we have $r_j\ge 0$.
\begin{Lemma} \label{k=3kiyakulemma}
In the setting above, we have $r_0=r_1=\cdots=r_{N-1}=0$.
\end{Lemma}
\textbf{Proof of lemma \ref{k=3kiyakulemma}}\;\;
Because every subscript $n$ is cyclic for $\tau_n^t$, it is enough to prove $r_j=0$ for only one specific $j$: e.g., $j=N-1$.
Let us take a specific initial condition from here on only in this proof:
\[
I_n^0=1\ (1\le n\le N),\ V_n^0=1\ (1\le n\le N-1),\ V_N^0=\frac{1}{x}.
\]
Then we have
\[
a_n=1,b_n=1\ (0\le n\le N),\  \lambda=K=1,\  \mu=1/x,
\]
using equations from \eqref{K} to \eqref{taun1}.
We have from equation \eqref{cn} that
\[
c_k=(k+1)x+N-k-1\ (k=0,1,\cdots, N-1).
\]
Using the equation \eqref{periodtau3kai} for $t=2$, we have
\begin{equation} \label{dN-1}
d_{N-1}=b_N\left[\frac{\lambda^2}{\mu} \sum_{k=0}^{N-1}\frac{(c_k)^2}{b_k b_{k+1}}\right]=x\sum_{k=0}^{N-1} (c_k)^2.
\end{equation}
If $x=-(N-k-1)/(k+1)$ then, we have $c_k=0$. Since $x\neq 0$ for $0\le k \le N-2$, we have $d_{N-1}\neq 0$ from \eqref{dN-1}.
Therefore we have proved that $d_{N-1}$ does not have a positive power of $c_k$ $(0\le k\le N-2)$ as a factor.
Thus $r_k=0$ $(0\le k\le N-2)$.

Lastly we prove that $d_{N-1}$ does not have a factor $c_{N-1}$.
By a cyclic permutation, it is enough to prove that $d_0$ does not have a factor $c_0$.
We have from \eqref{periodtau3kai} that
\[
d_0=x (c_0)^2 +\sum_{k=1}^{N-1} (c_k)^2.
\]
When we substitute $x=1-N$, $c_0=0$ in $d_0$, we have
\[
d_0=0 + \sum_{k=1}^{N-1} k^2 N^2=\frac{1}{6} N^3 (N-1)(2N-1)\neq 0.
\]
Thus $d_0$  does not have a positive power of $c_0$ as a factor. Therefore $r_{N-1}=0$.
\qed

Summing up these results, we have proved that
$d_j$ is irreducible for all $0\le j\le N-1$.
The co-primeness of distinct $d_j$ and $d_k$ $(j\neq k)$ follows immediately.
Finally we can prove that $d_j$ and $c_k$ are co-prime for $0\le j,k\le N-1$, since they are both irreducible and have different degrees.
Thus we have proved (B) for $k=3$.
\qed
\paragraph{The case of $\boldsymbol{k=4}$:}
Let us apply lemma \ref{locallemma} in the case of $m=2N$,
\begin{eqnarray*}
\{p_1, \cdots,p_m\}&=&\{b_1,\cdots, b_N,c_1,\cdots,c_N\},\\
\{q_1, \cdots, q_m\}&=&\{a_1,\cdots, a_N, b_1,\cdots, b_N\}.
\end{eqnarray*}
In the same manner as in the previous paragraph for $k=3$ we have the decomposition
\[
e_j=c_0^{s_0} c_1^{s_1} \cdots c_{N-1}^{s_{N-1}} \cdot H,
\]
where $H$ is an irreducible Laurent polynomial in $\mathbb{Z}[\{a_i^{\pm}\},\{b_i^{\pm}\}]$, and each $s_j\in\mathbb{Z}$, $s_j>0$.
Let us prove that $s_0=s_1=\cdots=s_{N-1}=0$, in order to prove that $e_j$ is an irreducible Laurent polynomial.
For this purpose we prove the following lemma.
\begin{Lemma} \label{eN-1}
The term $e_{N-1}$ does not have a positive power of $c_j$ as a factor for any $0\le j\le N-1$.
\end{Lemma}
\textbf{Proof of lemma \ref{eN-1}}\;\; 
Let us choose the same specific initial condition in this proof as in the previous paragraph
\[
a_n=1,b_n=1\ (0\le n\le N),\  \lambda=K=1,\  \mu=1/x,
\]
\[
c_k=(k+1)x+N-k-1\ (k=0,1,\cdots, N-1).
\]
\subparagraph{The case of $\boldsymbol{c_0}$:}
From the evolution of discrete Toda equation \eqref{dtoda5.5},
\[
c_{N-1} e_{N-1} = c_{N} e_{N-2} +\left( 1-\frac{\lambda^2}{\mu} \right)(d_{N-1})^2.
\]
From the induction hypothesis that $\{c_i\},\{d_i\}$ are coprime, we conclude that $e_{N-1}$
does not have a positive power of $c_{N}=c_0$ as a factor.
\subparagraph{The case of $\boldsymbol{c_{N-1}}$:}
By a cyclic permutation, it is enough to prove that $e_0$ does not have a factor $c_0$.
Equation \eqref{periodtau3kai} tells us that
\[
e_0=c_1\left[x\frac{d_0^2}{c_0 c_1}+\sum_{k=1}^{N-1} \frac{d_k^2}{c_k c_{k+1}}\right].
\]
In the case of $x=1-N$, we have $c_k=-kN$ $(k=0,1,\cdots, N-1)$, and
\[
d_0=\frac{1}{6}N^3(N-1)(2N-1)\neq 0.
\]
Therefore $e_0$ diverges if we take the limit $c_0\to 0$.
Thus $e_0$ cannot have a positive power of $c_0$ as a factor.
\subparagraph{The case of $\boldsymbol{c_i}$ $\boldsymbol{(1\le i\le N-2)}$:}
The equation \eqref{periodtau3kai} for $t=3$ and $c_{k+1}-c_{k}=x-1$ shows that
\begin{eqnarray}
e_{N-1}&=&c_N x \sum_{k=0}^{N-1} \frac{(d_k)^2}{c_k c_{k+1}} \notag \\
&=&c_N x \left[ \sum_{k=0}^{N-1} \frac{1}{c_{k+1}-c_{k}}\left( \frac{d_k^2}{c_k}-\frac{d_k^2}{c_{k+1}} \right) \right] \notag \\
&=&c_N \frac{x}{x-1}\left[ \frac{d_0^2}{c_0}-\frac{d_{N-1}^2}{c_N}+\sum_{k=1}^{N-1} \frac{d_k^2-d_{k-1}^2}{c_k} \right]. \label{eN-1keisan}
\end{eqnarray}
We have
\[
d_k=x \sum_{i=0}^k (c_i)^2 + \sum_{i=k+1}^{N-1} (c_i)^2.
\]
Thus $d_k-d_{k-1}=(x-1) c_k^2$. Therefore
the term $\frac{1}{c_k}$ in equation \eqref{eN-1keisan} is eliminated:
\begin{equation} \label{eN-1keisan2}
e_{N-1}=\frac{c_N x}{x-1}\left[ \frac{d_0^2}{c_0}-\frac{d_{N-1}^2}{c_N}+\sum_{k=1}^{N-1} (x-1) c_k (d_k+d_{k-1}) \right].
\end{equation}
Let us substitute $c_j=0$ $(x=1-\frac{N}{j+1})$ in the equation \eqref{eN-1keisan2} to obtain the following result:
\begin{equation}
e_{N-1}=-\frac{x N (N-1)}{180(x-1)} \cdot F,
\end{equation}
where
\begin{align}
F&=180j^4 + (390-420N) j^3+30(3N-2)(4N-5) j^2 \notag \\
&-2(2N-1)(34N^2-84N+47)j + 5(N-1)(N-2)(2N-1)^2. \label{valueF}
\end{align}
Derivation of \eqref{valueF} is in the appendix.
We have the following lemma on the positivity of $F$:
\begin{Lemma} \label{positiveF}
We have $F>0$ for all $N\ge 3$ and for all $1\le j\le N-2$.
\end{Lemma}
Proof of this lemma is straightforward but technical, therefore explained in the appendix.
From lemma \ref{positiveF}, we conclude that $e_{N-1}$ does not have a factor $c_j$ for $1\le j\le N-2$.
Summing up the three sub-paragraphs, we have proved lemma \ref{eN-1}.
\qed
\paragraph{The case of $\boldsymbol{k=5}$:}
Let us apply lemma \ref{locallemma} in the case of $m=2N$,
\begin{eqnarray*}
\{p_1, \cdots,p_m\}&=&\{d_1,\cdots, d_N,e_1,\cdots,e_N\},\\
\{q_1, \cdots, q_m\}&=&\{a_1,\cdots, a_N, b_1,\cdots, b_N\}.
\end{eqnarray*}
We can prove that these variables satisfy the conditions of lemma \ref{locallemma} from the induction hypotheses.
In the same manner as in the paragraph for $k=3$ we have the decomposition
\begin{equation} \label{fdecomp1}
f_j:= \tau_j^5 = d_0^{s_0} d_1^{s_1} \cdots d_{N-1}^{s_{N-1}}\cdot e_0^{t_0} e_1^{t_1}\cdots e_{N-1}^{t_{N-1}} \cdot H_1,
\end{equation}
where $s_i,t_i\in\mathbb{Z}_{\ge 0}$, and $H_1$ is an irreducible Laurent polynomial in the initial variables $\{a_i\}$, $\{b_i\}$.
Similarly, we also have another decomposition of $f_j$ as
\begin{equation} \label{fdecomp2}
f_j=c_0^{r_0} c_1^{r_1} \cdots c_{N-1}^{r_{N-1}}\cdot H_2,
\end{equation}
where $r_i\in\mathbb{Z}_{\ge 0}$, and $H_2$ is an irreducible Laurent polynomial in the initial variables.
Let us suppose that $f_j$ is not irreducible.
Since arbitrary two elements from $\{c_i\} \cup \{d_i\} \cup \{e_i\}$ are coprime,
the only possible decomposition of $f_j$ compatible with both \eqref{fdecomp1} and \eqref{fdecomp2} are one of the following two types:
\begin{equation}
f_j=Mc_k d_l, \label{fcd}
\end{equation}
or
\begin{equation}
f_j=M c_k e_l, \label{fce}
\end{equation}
where $M$ is a monomial in the initial variables $\{a_i\},\{b_i\}$.
Let us choose the same specific initial condition as in the previous paragraph
\[
a_n=1,b_n=1\ (0\le n\le N),\  \lambda=K=1,\  \mu=1/x,
\]
and take the limit $x\to 1$.
Then we have
\[
c_j=N,\ d_j=N^3,\ e_j=N^6,\ f_j=N^{10},
\]
for all $j\ge 0$.
We also have that the monomial $M\to \pm 1$.
Therefore
the degree (w.r.t. $N$) of left hand side of the equation \eqref{fcd} is $10$, while that of right hand side is $4$, which is a contradiction.
The degree of equation \eqref{fce} also has the same contradiction.
Thus $f_j$ does not have a decomposition, and is therefore irreducible.
Co-primeness of two terms is directly proved by the irreducibility.
\paragraph{The case of $\boldsymbol{k=6}$:}
The proof is just the same as in the case of $k=5$.
We note that $g_j=N^{15}$ under the same conditions as in the previous case.
\paragraph{The case of $\boldsymbol{k\ge 7}$:}
We have the following three types of decompositions at the same time
for $\tau_j^t$ $(t\ge 7)$:
\begin{align*}
\tau_j^t&=c_0^{r_0}\cdots c_{N-1}^{r_{N-1}}H_1=d_0^{s_0}\cdots d_{N-1}^{s_{N-1}}e_0^{t_0}\cdots e_{N-1}^{t_{N-1}} H_2\\
&=f_0^{p_0}\cdots f_{N-1}^{p_{N-1}} g_0^{q_0}\cdots g_{N-1}^{q_{N-1}}H_3,
\end{align*}
where $H_1,H_2,H_3$ are irreducible Laurent polynomials of initial variables.
Since from $c_i$ through $g_i$ are all irreducible elements, arbitrary two of which are coprime, we conclude that $r_i=s_i=t_i=p_i=q_i=0$ for all $i$.
Therefore $\tau_j^t$ is irreducible for $t\ge 7$. Co-primeness of $\tau_j^t$ and $\tau_i^t$ is proved by the irreducibility of themselves and the cyclic property in terms of the subscripts. Co-primeness of $\tau_j^t$ with arbitrary $\tau_j^s$ with $s< t$ is proved by the irreducibility of $\tau_j^t$ and by the fact that $\tau_j^t$ and $\tau_k^s$ has  different degrees if $t\neq s$.
The proof of proposition \ref{kiyakulemmaperiod} is finished.
\qed
The proof of theorem \ref{periodthm} is now completed.\qed

Remember that theorem \ref{periodthm} is for the transformed function $\tilde{\tau}_n^t$, and the statement for the original $\tau_n^t$ is as follows:
\begin{Corollary} \label{periodthm2}
The function $\tau_n^t$ is an irreducible Laurent polynomial of the initial variables and a power of $(\lambda^2/\mu -1)$:
\[
\tau_n^t\in \mathbb{Z}[(\lambda^2/\mu -1)^{\pm},(\tau_n^0)^{\pm},(\tau_n^1)^{\pm};0\le n\le N-1],
\]
and two distinct terms are co-prime.
\end{Corollary}
Using corollary \ref{periodthm2}, we can prove our main theorem of co-primeness of the discrete Toda equation with periodic boundary condition.
\begin{Theorem}
Let us take $N\ge 6$.
The solution $I_n^t$, $V_n^t$ of the periodic discrete Toda equation (\eqref{dtodaIV1} and \eqref{dtodaIV2} with $I_{n+N}^t=I_n^t$ and $V_{n+N}^t=V_n^t$)  satisfies the following `co-prime' property:
Let us define the set $\mathcal{D}=\{I_n^t\}_{0\le n\le N-1, 0\le t}\cup\{V_n^t\}_{0\le n\le N-1, 0\le t}$.
Two elements $D_n^t$ and $D_m^s$ in the set $\mathcal{D}$
do not have common factors other than monomials of the initial variables, on condition that $N-3\ge |n-m|\ge 3$ or $|t-s|\ge 2$, where $n,m,t,s$ can be considered as the values modulo $N$.
\end{Theorem}
\textbf{Proof}\;\;
We use the relation \eqref{tauIV}, and the co-primeness of $\tau_n^t$ and $\tau_m^s$ for $(n,t)\neq (m,s)$ in corollary \ref{periodthm2}. The factor $(1-\lambda^2/\mu)$ is eliminated in $I_n^t$ and $V_n^t$ from \eqref{tauIV}.
The rest of the proof is the same as in the previous theorem \ref{locallemma}.\qed
Note that we are not stating that no pair of two terms is co-prime when $N<6$. The above theorem is a sufficient condition (good enough for large system size $N$) for co-primeness under the periodic boundary condition.
\section{Concluding remarks and discussions}
In this paper, we studied the discrete Toda equation in terms of the properties of irreducibility and co-primeness of the solutions.
We studied the discrete Toda equation under three different cases of boundary conditions: semi-infinite, molecule and periodic. We proved the coprime condition for all the three cases.
Our results, along with preceding results for the discrete KdV equation and the Quispel-Roberts-Thompson type mappings \cite{dKdVSC,dKdVSC2}, justify our assertion that the coprime property is an integrability detector.
Since our results include the case of the equation with periodic boundary condition, which cannot be easily dealt with the singularity confinement approach, the coprime property is expected to be applicable to wider class of integrable and non-integrable mappings under various conditions than conventional integrability tests.

The co-primeness has another advantage that it contains global information on the common factors of the general terms of the equation.
Because of this global property, rigorously proving the co-primeness sometimes involves long and technical calculations. However, when we use the co-primeness as an aid to \textbf{conjecture} the integrability of the given equation, difficulty of a proof does not pose a problem. We just have to compute a finite number of terms using a mathematical software, and observe the appearance of common factors. If the computation is too heavy, it may be a good idea to substitute arbitrary \textbf{integer} numbers to some of the independent variables, which greatly reduces the computing time.
Indeed, we have to note that the irreducibility and co-primeness are not preserved after substituting numbers to the variables, but the result is usually practical enough to grasp the appearance of common factors.

One of the future works is to study the co-primeness of other discrete integrable and non-integrable equations. In particular, we will investigate the equations, for which several integrability criteria give conflicting results on their integrability.
For example the Hietarinta-Viallet equation \cite{HV} passes the singularity confinement test, but it has a positive algebraic entropy \cite{BV}, which is an indication of non-integrability.
Some of the linearizable discrete mappings \cite{RGSM} do not pass the singularity confinement test, although their algebraic entropy is zero.
By applying the co-prime criterion and by investigating the common factors even more closely, we expect to obtain convincing results on the integrability of these equations in future works.
It is also a good idea to investigate the relation of our results with other integrability criteria such as the $p$-adic number theoretic interpretation of the confined singularities \cite{KMTT}, and the singularity confinement for ultra-discrete systems \cite{Joshi}, which has recently been studied in relation to the tropical geometry \cite{Ormerod}.

\section*{Acknowledgments}
The authors wish to thank Prof. R. Willox and Dr. T. Mase for useful comments.
This work is partially supported by Grant-in-Aid for Scientific Research of JSPS ($26\cdot 242$).

%
%
%appendix from here
%
%

\appendix
\section{Appendix: On equation \eqref{valueF} and lemma \ref{positiveF}}
We first give a detail on how to derive equation \eqref{valueF}.
Value $F$ is obtained by substituting $c_j=0$ $(x=1-\frac{N}{j+1})$ in the equation \eqref{eN-1keisan2}.
If $c_j=0$, then $c_k=\frac{j-k}{j+1}N$. Therefore we have
\begin{align*}
d_k&=\left(1-\frac{N}{j+1}\right)\sum_{i=0}^k \left(\frac{j-i}{j+1} N \right)^2+\sum_{i=k+1}^{N-1} \left( \frac{j-i}{j+1}N\right)^2\\
&=\frac{N^2}{(j+1)^2}\left[ -\frac{N}{j+1}\sum_{i=0}^k (j-i)^2+\sum_{i=0}^{N-1} (j-i)^2 \right]\\
&=\frac{N^3}{6(j+1)^3}\Big[(1 + 7 j + 6 j^2 + 6 j^3 - k + 6 jk - 6 j^2 k \\
&- 3 k^2 + 6 j k^2 - 
  2 k^3 - 3 N - 9 j N - 6 j^2 N + 2 N^2 + 2 j N^2)\Big].
\end{align*}
The former two terms in the right hand side of equation \eqref{eN-1keisan2} is calculated as
\begin{align*}
&\frac{d_0^2}{c_0}-\frac{d_{N-1}^2}{c_0}=\frac{N^5(N-1)}{36j (j+1)^5}(6 j^2 - N - 6 j N + 2 N^2)\\
&\times (2 + 14 j + 18 j^2 + 12 j^3 - 7 N - 24 j N - 18 j^2 N + 7 N^2 + 
  10 j N^2 - 2 N^3).
\end{align*}
The last term in the right hand side of equation \eqref{eN-1keisan2} is calculated as
\begin{align*}
&\sum_{k=1}^{N-1} (x-1)c_k (d_k+d_{k-1})\\
&=\frac{1}{30(j+1)^5}(10 j N^5 + 70 j^2 N^5 + 90 j^3 N^5 + 60 j^4 N^5 - 6 N^6 - 80 j N^6\\
& - 
  225 j^2 N^6 - 210 j^3 N^6 
- 60 j^4 N^6 + 25 N^7 + 155 j N^7 + 
  245 j^2 N^7 \\
&+ 120 j^3 N^7 - 35 N^8
 - 115 j N^8 - 90 j^2 N^8 + 
  20 N^9 + 30 j N^9 - 4 N^{10}).
\end{align*}
Summing up these results we obtain the expression for $e_{N-1}$ and $F$ as in \eqref{valueF}.
\paragraph{Proof of lemma \ref{positiveF}:}
Next we give proof of lemma \ref{positiveF}.
For $N=3,4,5,6$, we can prove that $F$ does not factorize in $\mathbb{Z}[j]$ except for a constant factor:
\begin{equation*}
F=\left\{
\begin{array}{cl}
10(18j^4-87j^3+147j^2-101j+25) & (N=3)\\
30(6j^4-43j^3+110j^2-119j+49) & (N=4)\\
18(10j^4-95j^3+325j^2-477j+270) & (N=5)\\
2(90j^4-1065j^3+4560j^2-8437j+6050) & (N=6)
\end{array}.
\right.
\end{equation*}
Next we prove $F>0$ for $N\ge 7$ and $1\le m\le N-2$.
Let us take $x:=\frac{m}{N}$ and rewrite $F$
as
\[
\tilde{F}:=\frac{F}{N^4}=f_0(x)+\frac{1}{N}f_1(x)+\frac{1}{N^2}f_2(x)+\frac{1}{N^3}f_3(x)+\frac{10}{N^4},
\]
where
\begin{eqnarray*}
f_0(x)&=&180x^4-420x^3+360x^2-136x+20,\\
f_1(x)&=&2(13x-10)(15x^2-15x+4),\\
f_2(x)&=&300x^2-356x+105,\\
f_3(x)&=&94x-55.
\end{eqnarray*}
We prove $\tilde{F}>0$ for $0<x<1$.
We have
\[
\frac{1}{N^2}f_2(x)+\frac{1}{N^3}f_3(x)\ge -\frac{184N^2-232N+2209}{300N^4},
\]
and this minimum
is reached at $0<x=\frac{178N-47}{300N}<1$.
Since $\frac{-(184N^2-232N+2209)}{300N^4}$
is strictly increasing for $N\ge 7$, we have
\[
\frac{1}{N^2}f_2(x)+\frac{1}{N^3}f_3(x)\ge -\frac{9601}{720300} ,
\]
for all $N\ge 7$.
Therefore the proof is complete when we prove that
\[
\tilde{F}_N(x):=f_0(x)+\frac{1}{N}f_1(x)-\frac{9601}{720300}
\]
is positive for $0<x<1$ and $N\ge 7$. (Note that $\tilde{F}>\tilde{F}_N(x)$.)
The function $f_0(x)$ has the minimum at
\[
x=x_0:=\frac{1}{12}\left\{7+5^{-1/3}\left( 11-4\sqrt{6} \right)^{1/3}+5^{-1/3}\left( 11+4\sqrt{6} \right)^{1/3}\right\}=0.759\cdots,
\]
and the minimum is
\begin{align*}
f_0(x_0)&=\frac{1}{192}\Big\{ 113-2 \cdot 5^{1/3} \left((11+4\sqrt{6})^{2/3}+(11-4\sqrt{6})^{2/3}\right) \\
&-5^{-1/3}\left( (11+4\sqrt{6})^{4/3}+(11-4\sqrt{6})^{4/3} \right)\Big\}=0.255\cdots.
\end{align*}
Therefore if $f_1(x)\ge 0$ $(\frac{10}{13}\le x<1)$ then $\tilde{F}_N(x)>0.255-\frac{9601}{720300}>0$ for all $N$, and therefore the proof of lemma \ref{positiveF} is done.
In the case of $f_1(x)<0$ $(0<x<\frac{10}{13})$, we have
\[
\tilde{F}_{N+1}(x)-\tilde{F}_N(x)=-\frac{1}{N(N+1)}f_1(x)>0.
\]
Thus we only have to prove that $\tilde{F}_7(x)>0$ for $0<x<\frac{10}{13}$.
The function $\tilde{F}_7(x)$ attains its local minima at two points
\begin{eqnarray*}
x&=&\alpha=\frac{1}{168}\left[ 85-\frac{1-\sqrt{-3}}{2}\left( \rho_{+} \right)^{1/3} -\frac{1+\sqrt{-3}}{2}\left( \rho_{-} \right)^{1/3} \right],\\
x&=&\beta=\frac{1}{168}\left[85+\left( \rho_{+} \right)^{1/3}+\left( \rho_{-} \right)^{1/3}\right],
\end{eqnarray*}
where $\rho_{\pm}=(4121 \pm 56\sqrt{-478469})/5$. Note that $(a+bi)^{1/3}+(a-bi)^{1/3}\in\mathbb{R}$ for $a,b\in\mathbb{R}$, $i=\sqrt{-1}$.
Numerical calculation shows that
\[
\alpha=0.305\cdots,\; \beta=0.714\cdots,
\]
and
\[
\tilde{F}_7(\alpha)=0.222\cdots>0,\; \tilde{F}_7(\beta)=0.119\cdots>0.
\]
\qed
%%%%%%%%%%%%%%%%%%%%%%%%%%%%%
%
%
%

%

\end{document}